\documentclass{appolb}
\usepackage{epsfig}
\def \pao {~[Pierre Auger Collaboration]}

\begin{document}

% \eqsec  % uncomment this line to get equations numbered by (sec.num)
\title{Status and Recent Results from the Pierre Auger Observatory%
\thanks{Presented at the XXXVII International Symposium on Multiparticle Dynamics, Berkely, 2007}%
% you can use '\\' to break lines
}
\author{J. R. T. de Mello Neto$^1$, for the Pierre Auger Collaboration$^2$ % Put here the name(s) of the Author(s)
\address{$^1$Instituto de F\'isica, Universidade Federal do Rio de Janeiro, Rio de Janeiro, RJ, and
         Department of Astronomy and Astrophysics, The University of Chicago, Chicago, IL 60637  \\
$^2$Observatorio Pierre Auger, Av. San Mart\'in Norte 304, (5613),Malarg\"ue, Mendoza, Argentina
%\and
%the Name(s) of other Author(s)
%\address{and their affiliation}
}
}

\maketitle
\begin{abstract}
We present the status and the recent measurements from the Pierre Auger Observatory. The energy
spectrum will be described and its steepening discussed.   The mass composition is addressed with
the  measurements of the variation of the
depth of shower maximum with energy. We also report on upper limits in the primary photon fraction.
And finally, searches for anisotropies of cosmic rays arrival directions are reported.

\end{abstract}

% observatory, hybrid technique
% spectrum
% composition
% hadronic models
% anisotropy - list of unconfirmed results and correlation with VC

\section{The Pierre Auger Observatory}
The Pierre Auger Observatory will be completed in early 2008 and will consist
 of 1600 10m$^2$ $\times$ 1.2\,m   water-cherenkov detectors  deployed over 3000
 km$^2$ on a 1500 m hexagonal grid overlooked by four fluorescence detectors (FD),
 each of them with 6 telescopes.  The surface detector (SD)  stations
sample at the ground level the charged particles in the shower front that 
cross the stations. The fluorescence telescopes  can record the ultraviolet 
light emitted as the shower
crosses the atmosphere. By September of 2007 about 90\% of the SD detector
and all 24 FD telescopes were operational. The accumulated exposure\footnote{Unless otherwise
 noted, the data shown here was recorded from January 2004 to the end of February 2007
and was shown at the 30$^{th}$ International Cosmic Ray Conference, in M\'erida,
 Mexico, July, 2007. }  up to 28 February
2007 was 5165 km$^2$~sr~yr, about the same as the monocular HiRes
 and three times larger then AGASA \cite{Watson_ICRC}.

\begin{figure}[htb]
\epsfig{file=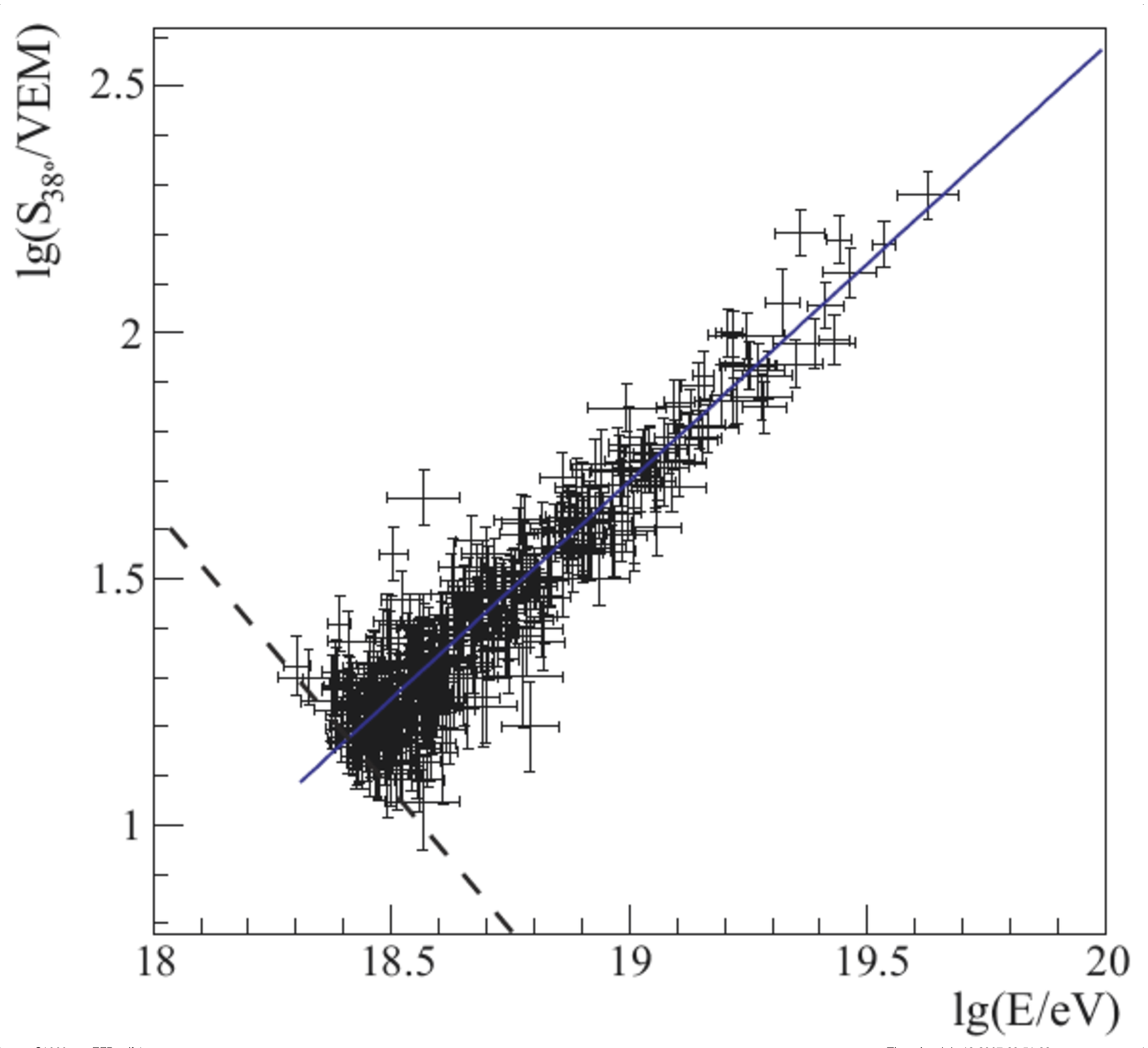, width=5.9cm} \epsfig{file=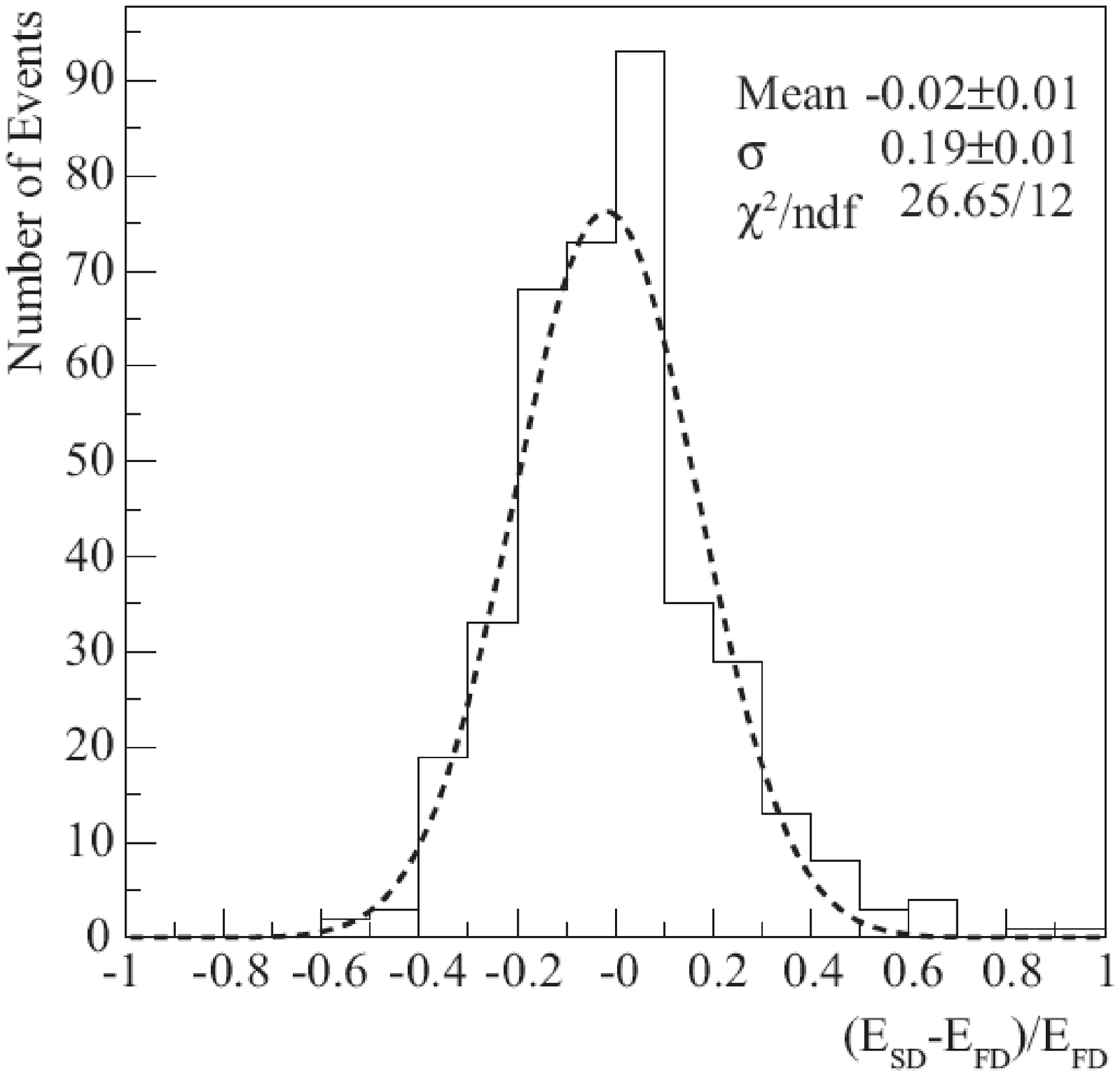, width=5.9cm}
\vspace*{-0.30cm}
\caption{\label{F-1}Left: correlation between the SD parameter $S_{38}$ (see text)
 and the reconstructed FD energy of Auger hybrid showers. Right: fractional dispersion of
 the SD/FD energy correlation.}
\end{figure}

The Pierre Auger Observatory was designed as a hybrid detector, i.e., to use
both fluorescence and ground array techniques in a extensive way. About 10\% of the events
are detected simultaneously by both techniques. From them two variables can be measured: the
ground parameter  $S_{38}$ and the energy as measured by the fluorescence detector. The 
correlation of the two variables is shown in fig. \ref{F-1} left. The dispersion is shown in 
the right.

%The measurement of
%the energy of the primary cosmic ray for events that are detected simultaneously
%by ground stations and fluorescence telescopes is done with a cross-calibration technique that
%correlates the ground parameter $S_{38}$ (the lateral distribution function of the
%energy deposition measured from the shower axis calculated at R=1000 m and normalized to $38^{\circ}$) with the calorimetric
%energy deposited in the atmosphere as the shower develops (see fig. \ref{F-1}). 
With
this calibration the energy of the showers that are detected by the SD only (tenfold more statistics)
can be estimated in a empirical~way~\cite{Roth_ICRC}.

% \footnote{The \emph{missing energy}
%of neutrinos and muons is estimated with the use of Monte Carlo, with tipical correction of the order
%of 12\% for E=10^{19} EeV} that contributes ~4\% to the total uncertainty in the FD energy (21\%)}

\section{Recent results}

\begin{figure}[htb]
\epsfig{file=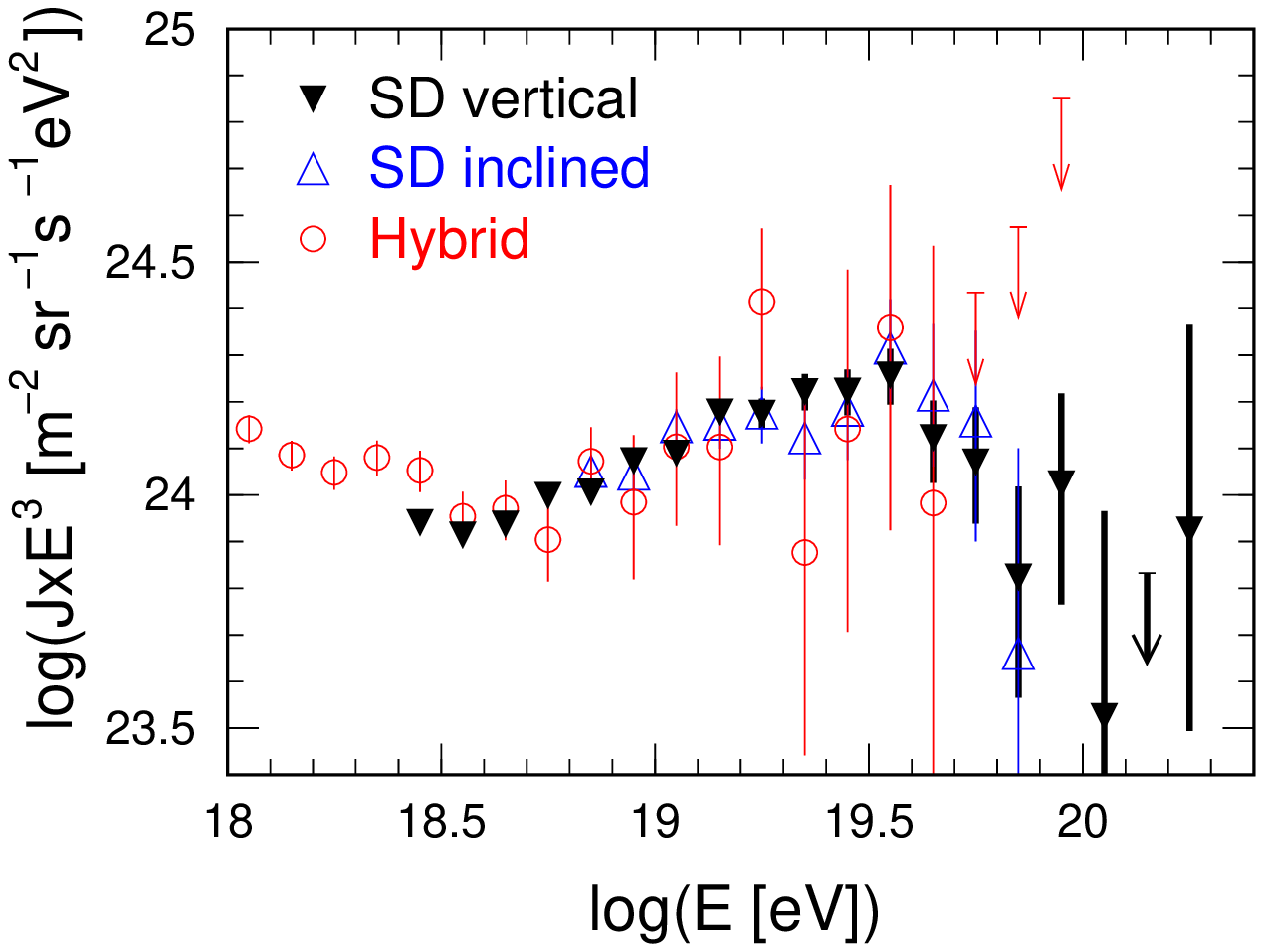, width=5.9cm} \epsfig{file=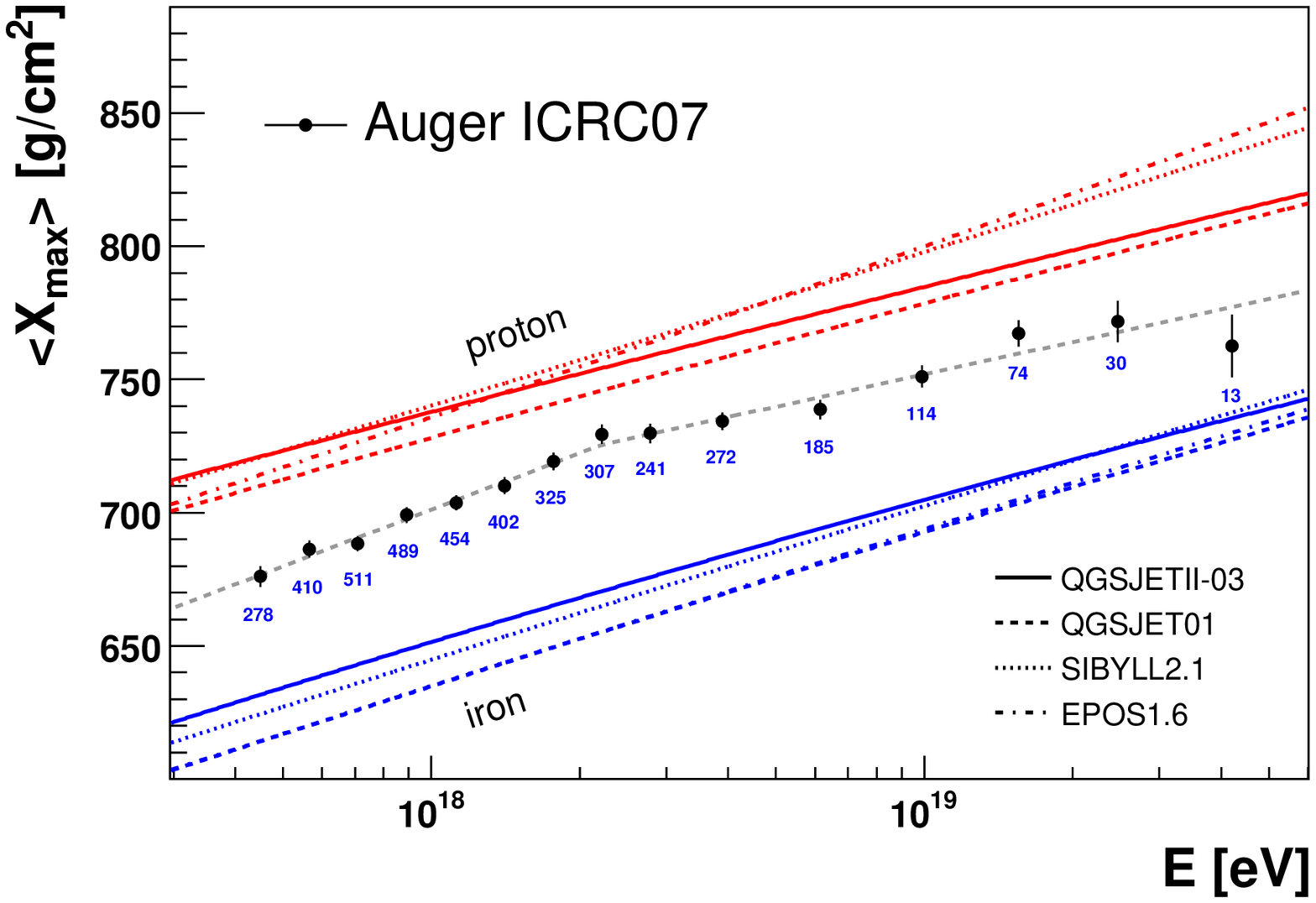, width=5.9cm}
\vspace*{-0.30cm}
\caption{\label{F-2}Left: The energy spectrum multiplied by $E^3$ derived from SD using showers
at zenith angles above (filled triangles) and below (opened triangles) $60^{\circ}$, together
with the spectrum derived from the hybrid data set (red circles) \cite{Yamamoto_ICRC}.
Right: The depth of maximum as a function of energy. The upper sets of lines are predictions
of protons for several models. the lower set are under the assumption of Fe nuclei \cite{Unger_ICRC}}
\end{figure}

In fig. \ref{F-2} left, the energy spectrum multiplied by $E^3$ from SD data using showers at
 zenith angle above and below 60$^{\circ}$, together with the spectrum obtained from the hybrid
data set are shown. The three spectra obtained using different methods agree well and a combined fit is performed. Only statistical errors are quoted.    A flattening of
the slope of the energy spectrum from
 $(-3.30 \pm 0.06)$ to $(-2.62 \pm 0.03)$ is observed at $ 10^{18.6}$ eV and above
 $10^{19.6}$ eV the spectrum gets steeper, with slope $(-4.14 \pm 0.42)$. The number of
events expected if
the power-law observed at  $ 10^{18.6}$ eV were extended above  $10^{19.6}$ eV and $10^{20}$ eV are
$132\pm9$ and $30\pm2.5$ whereas we observe only 58 events and 2 events, respectively \cite{Yamamoto_ICRC}.
So a sharp suppression in the last decade of the UHECR spectrum is observed
(as it was already shown by the HiRes experiment ~\cite{HiResAbbasiEtAl2007}).
Whether this is due to the expected GZK cuttoff or if it is due to other phenomena like a limit on the
acceleration process, is an open question and needs further investigation. The ``ankle'' is
determined to be at $E_{ankle}=10^{18.65 \pm 0.04} $ eV.

The depth of shower maximum depends on the primary mass and comparisons with Monte Carlo models
can be made to extract information about the mass composition of the UHECR. Hybrid events allow the
measurement of $X_{max}$ and have the geometry of the shower well constrained by the SD information.
The data is shown in figure  \ref{F-2} right. After appropriate quality and uniformity
cuts, the $X_{max}$ resolution is less than 40 g/cm$^2$.    A single slope does not seem to fit the data
and the rate of increase of $X_{max}$ with energy is smaller for the region above
$2 \times 10^{18}$ eV than below. More data will be necessary to establish the tendency to 
heavier composition at high energies.

Primary photon showers are easier to distinguish from primary hadron showers, since they penetrate deeper in the
atmosphere. This results in a larger rise time of the signal produced in the SD stations and in
 a smaller radius of curvature of the shower front.  Those two variables can be measured precisely
in the surface detector and are combined in one single variable through a principal component analysis.
No event up to now was identified with a photon. The absence (or very small fraction) of photons in the
incident primary cosmic ray flux allows stringent photon limits to be set up. For instance, at 10 EeV
the upper limit \cite{photon_lim} in the integral flux fraction produced by primary photons is about 2\%.  This sets
 constraints to \emph{top-down models}, where the UHECR are the decay products of heavy primordial
 particles or produced by topological defects.

%%\

One of the major aims for the experiment is the measurement of the anisotropy of the arrival direction
 of the UHECR. One expects that protons with energy of about 50 EeV or higher to deviate two or three
degrees
in the galactic magnetic field. In order to calculate the significance of the results in a unbiased way,
an \emph{a priori} specification of the details of the analysis to be done with \emph{future} data
is required. Any hint of a discovery found in a exploratory analysis will be subjected to a
 \emph{prescription} that requires new data. Previous claims of anisotropy from the Galactic center and of
clustering at high energies, as well as correlation with BL-Lacs were not confirmed by Auger\cite{anisot_ICRC}.

\section{Summary and Perspectives}
There are enhancements already approved and funded for the South Observatory. HEAT will
reduce the energy threshold down
to 0.1 EeV
and AMIGA will detect showers down to 0.1 EeV and measure their muon content \cite{Etchegoyen_ICRC}.
This will allow precise measurements  of
the shower composition in the ``ankle'' region and constrain hadronic models in this energy range.   % \cite{Tanco_ICRC}.
 The North Observatory,
planned to be built in Colorado, USA, possibly with a higher threshold in the detector energy, will provide  full sky coverage with enhanced statistics at the
end of the cosmic ray spectrum.

\end{document}